\documentclass[%
 aip,
cp,  
 amsmath,amssymb,
 reprint,%
]{revtex4-2}

\usepackage{amsmath}
\usepackage{graphicx}
\usepackage{dcolumn}
\usepackage{bm}
\usepackage{physics}
\usepackage[utf8]{inputenc}
\usepackage[T1]{fontenc}
\usepackage{mathptmx} 

\begin{document}

\title{Including Tetraquark Operators in the Low-Lying Scalar Meson Sectors in Lattice QCD}

\author{Daniel Darvish} 
 \email[Corresponding author: ]{ddarvish@andrew.cmu.edu}
  \affiliation{Department of Physics, Carnegie Mellon University, Pittsburgh, PA 15213, USA}
\author{Ruairí Brett}
 \affiliation{Department of Physics, The George Washington University, Washington, DC 20052, USA}
\author{John Bulava}
  \affiliation{CP3-Origins, University of Southern Denmark, Campusvej 55, 5230 Odense M, Denmark}
\author{Jacob Fallica}
 \affiliation{Department of Physics and Astronomy, University of Kentucky, Lexington, KY 40506, USA}
\author{Andrew~Hanlon}
 \affiliation{Helmholtz-Institut Mainz, Johannes Gutenberg-Universit\"at, D 55099 Mainz, Germany}
\author{Ben Hörz}
  \affiliation{Nuclear Science Division, Lawrence Berkeley National Laboratory, Berkeley, CA 94720, USA}
\author{Colin Morningstar}
  \affiliation{Department of Physics, Carnegie Mellon University, Pittsburgh, PA 15213, USA}

\date{\today} 

\begin{abstract}
Lattice QCD allows us to probe the low-lying hadron spectrum in finite-volume using a basis of single- and multi-hadron interpolating operators. Here we examine the effect of including tetraquark operators on the spectrum in the scalar meson sectors containing the $K_0^*(700)$ ($\kappa$) and the $a_0(980)$ in $N_f = 2 + 1$ QCD, with $m_\pi \approx 230$ MeV. Preliminary results of  additional finite-volume states found using tetraquark operators are shown, and possible implications of these states are discussed.
\end{abstract}

\maketitle

\section{Introduction}
It has been suggested that the light scalar mesons $K_0^*(700)$ (here referred to as the $\kappa$) and $a_0(980)$ could have tetraquark content~\cite{Jaffe:2004ph, AMSLER200461, Close_2002, PhysRevLett.93.212002}. To date, there have been a small number of studies investigating tetraquarks on the lattice using light quarks. In 2010, Prelovsek et al. investigated the $\sigma$ and $\kappa$ as possible tetraquark candidates, but neglected disconnected diagrams in their calculations~\cite{Prelovsek2010}. Using tetraquark interpolators, they found an additional light state in both the $\sigma$ and $\kappa$ channels. In 2013, the ETM collaboration examined the $a_0(980)$ and $\kappa$ using four-quark operators~\cite{Alexandrou2013}, though they also neglected disconnected diagrams in their calculations. They found no evidence of an additional state that can be interpreted as a tetraquark. In 2018, Alexandrou et al. conducted a study of the $a_0(980)$ with four-quark operators~\cite{Alexandrou2018}, including disconnected contributions. In their study, they found an additional finite-volume state in the sector containing the $a_0(980)$ meson, which couples to a diquark-antidiquark interpolating field. Additionally, they conclude that disconnected diagrams have drastic effects on their results, and thus cannot be neglected.

Here we investigate the possible role of tetraquark operators in lattice QCD in the symmetry channels of the $\kappa$ ($I = \frac{1}{2}$, $S=1$, $P = +1$, $J=0$) and $a_0(980)$ ($I=1$, $S=0$, $P=+1$, $G=-1$, $J=0$). We perform Monte Carlo calculations using $412$ gauge field configurations on an anisotropic ($\frac{a_s}{a_t} \approx 3.451$) lattice of size $32^3\times 256$, with a length of $3.74$ fm and a pion mass of approximately $230$ MeV. We extract two spectra in each symmetry channel: one using a basis of only single- and two-meson operators, and one using a basis that also includes a tetraquark operator selected from hundreds of tetraquark operators which were tested. We find that including a tetraquark operator yields an additional finite-volume state in each symmetry channel. In this work, we use the stochastic LapH method~\cite{slaph} to evaluate all diagrams in our calculations, including \textit{all} disconnected contributions.
\section{Spectroscopy in lattice QCD}
We obtain our finite-volume QCD spectra by calculating discretized path integrals in Euclidean spacetime. Consider an operator $\overline{\mathcal O}(t)$ that acts on the vacuum state $\ket{0}$, creating a state at time $t$, and a corresponding operator $\mathcal O(t)$ that annihilates such a state at time $t$. In imaginary time, we can in principle extract the energy spectrum from the following time-ordered correlation function,
\begin{equation}\label{eq:correlator}
  \bra{0}T\, \mathcal O(t + t_0) \overline{\mathcal O}(t_0)\ket{0} = \sum_n \bra{0}\mathcal O(0) \ket{n}\bra{n} \overline{\mathcal{O}}(0)\ket{0} e^{-E_n t},
\end{equation}
where $\ket{n}$ is the $n^{\mathrm{th}}$ ordered energy eigenstate of the theory corresponding to energy $E_n$, and we have shifted the vacuum energy $E_0$ to be zero. (Henceforth, $E_0$ will refer to the \textit{ground state} energy in each symmetry sector.)

Expectation values in Euclidean spacetime can be calculated using a path integral over all quark fields $\psi$, $\overline \psi$, and all link variables representing the gauge fields, $U$:
\begin{equation}
  \begin{aligned}
    \langle\mathcal O\rangle &= \frac{1}{Z} \int \mathcal D\left[\overline \psi, \psi, U \right]\, \mathcal O\, e^{-S\left[\overline \psi, \psi, U \right]},\\
    Z &= \int \mathcal D\left[\overline \psi, \psi, U \right]\, e^{-S\left[\overline \psi, \psi, U \right]}.
  \end{aligned}
\end{equation}
In order to evaluate these integrals on the lattice, we discretize spacetime in the above integral and carefully design operators that create states having nonzero overlap with the states of interest. We \textit{analytically} integrate out the quark fields, and use Monte Carlo methods to integrate over the link variables, using gauge field configurations generated by the Hadron Spectrum Collaboration~\cite{PhysRevD.78.054501, PhysRevD.79.034505, CLARK2005835}.

By calculating Eq. (\ref{eq:correlator}) on the lattice, we can, in practice, only fit the lowest one or two energies. At small $t$, there is a high signal-to-noise ratio in the Monte Carlo determination of the correlator, but there are many excited states that contribute. It would be better if we could devise a way to extract more than just the lowest one or two states. In order to do this, we construct an $N\times N$ \textit{matrix} of correlators,
\begin{equation}\label{eq:cij}
  \mathcal C_{i j}(t) \equiv\bra{0}T \mathcal{O}_{i}\left(t+t_{0}\right) \overline{\mathcal{O}}_{j}(t)\ket{0},
\end{equation}
using a large basis of single- and multi-hadron interpolating operators. As a preliminary step, we rescale Eq. (\ref{eq:cij}) in order to compensate for varying normalizations between the different operators:
\begin{equation}
  C_{i j}(t) \equiv \frac{\mathcal{C}_{i j}(t)}{\sqrt{\mathcal{C}_{i i}\left(\tau_{N}\right) \mathcal{C}_{j j}\left(\tau_{N}\right)}},
\end{equation}
where the \textit{normalization time} $\tau_N$ is taken at some early time, e.g. $\tau_N < 4$. We will also assume that for sufficiently large $t$, and for sufficiently small $Z_i^{(n)} \equiv \bra{0}\mathcal O_i \ket{n}$, we can well approximate the correlation matrix by,
\begin{equation}
  C_{ij}(t) \approx \sum_{n=0}^{N-1} Z_{i}^{(n)} Z_{j}^{(n) *} e^{-E_{n} t}.
\end{equation}
With our truncated $C(t)$, we then solve the following generalized eigenvalue problem:
\begin{equation}
  C(t) v_{n}\left(t, \tau_{0}\right)=\lambda_{n}\left(t, \tau_{0}\right) C\left(\tau_{0}\right) v_{n}\left(t, \tau_{0}\right), \quad t>\tau_{0}.
\end{equation}
It is shown in Ref.~\cite{LUSCHER1990222} that for $\tau_0 \geq t/2$,
\begin{equation}
  \lambda_{n}\left(t, \tau_{0}\right)=e^{-E_{n}\left(t-\tau_{0}\right)}\left(1+\mathcal{O}\left(e^{-\left(E_{N}-E_{n}\right) t}\right)\right).
\end{equation}
(In practice, we find that the condition $\tau_0 \geq t/2$ is not strictly necessary.) We can therefore fit the lowest $N$ energies of the spectrum by fitting $\lambda_n(t)$ to single- or two-exponentials. In practice, the excited state contamination in $\lambda(t,\tau_0)$ scales such that it is prudent to discard the highest few levels, or rather, to use more operators in the basis than levels we wish to determine.
\section{Operator construction}
In our operator bases, we include single- and two- meson operators, as well as tetraquark operators. We make use of smeared, gauge-covariantly displaced quark fields, and stout-smeared link variables (introduced in Ref.~\cite{PhysRevD.69.054501}) in our operator construction. For example, an elemental meson operator of definite-momentum $\boldsymbol{p}$ at time $t$ can be written as follows,
\begin{equation}\label{eq:meson_op}
  \Phi_{\alpha \beta ; i j k}^{A B}(\boldsymbol{p}, t)=\sum_{\boldsymbol{x}} e^{-i \boldsymbol{p} \cdot\left(\boldsymbol{x}+\frac{1}{2}\left(\boldsymbol{d}_{\alpha}+\boldsymbol{d}_{\beta}\right)\right)} \delta_{a b} \overline{q}_{a \alpha i}^{A}(\boldsymbol{x}, t) q_{b \beta j k}^{B}(\boldsymbol{x}, t),
\end{equation}
where capital Latin indices denote flavor, lowercase Latin $a$ and $b$ denote color, Greek indices are Dirac spin, and lowercase Latin $i$, $j$, $k$, denote quark displacement. The vectors $\boldsymbol{d}_\alpha$ and $\boldsymbol{d}_\beta$ are quark displacement vectors, and are present to ensure proper transformation under $G$-parity. To form the final operators out of our elemental operators, we project the elemental operators onto various symmetry channels according to isospin, parity, $G$-parity, octahedral little group, etc. That is, to form a meson operator $M_{l}(t)$ that transforms irreducibly under all symmetries of interest (labeled by the compound index $l$) at time $t$, we must 
take a linear combination of our elemental meson operators, $M_{l}(t)=c_{\alpha \beta}^{(l)} \Phi_{\alpha \beta}^{A B}(\boldsymbol{p}, t)$. To form a two-meson operator $\mathcal{O}_l(t)$, we would follow a similar procedure and project the product of two final meson operators $M^{a}_{l_a}(t) M^{b}_{l_a}(t)$ onto a final symmetry channel $l$: $\mathcal{O}_l(t) = c^{(l)}_{l_a l_b} M^{a}_{l_a}(t) M^{b}_{l_a}(t)$.

In order to construct a tetraquark operator, we must consider the various ways to construct a color-singlet four-quark object out of four quark fields. As seen in Ref.~\cite{pittir33243}, the Clebsch-Gordon decompositions show that the only way to construct a color-singlet is by using two quarks and two antiquarks, and that doing so yields two linearly independent color singlet objects:
\begin{equation}
\begin{array}{l}
    {3 \otimes 3 \otimes 3 \otimes 3=3\oplus3\oplus3\oplus\overline{6}\oplus\overline{6}\oplus15\oplus15\oplus15\oplus15},\\
    {3 \otimes 3 \otimes 3 \otimes \overline{3}=\overline{3}\oplus\overline{3}\oplus\overline{3}\oplus6\oplus6\oplus6\oplus\overline{15}\oplus\overline{15}\oplus24},\\
    {3 \otimes 3 \otimes \overline{3} \otimes \overline{3}=1\oplus1\oplus8\oplus8\oplus8\oplus8\oplus10\oplus\overline{10}\oplus27}.
\end{array}
\end{equation}
There are 81 basis vectors formed by the quark fields, $p_{\alpha}^{*}(x) q_{\beta}^{*}(x) r_{\gamma}(x) s_{\mu}(x)$, where each $r$, $s$ transforms as a color vector in the fundamental $3$ irrep, and so, $p^{*}$, $q^{*}$ transform in the $\overline 3$ irrep. We need two linearly independent and gauge-invariant combinations of these to exhaust all possible tetraquark operators. It is easy to see that the following combinations are both linearly independent and gauge-invariant (and thus form our elemental tetraquark operators):
\begin{equation}\label{eq:tsta}
\begin{aligned} T_{S} &=\left(\delta_{\alpha \gamma} \delta_{\beta \mu}+\delta_{\alpha \mu} \delta_{\beta \gamma}\right) p_{\alpha}^{*}(x) q_{\beta}^{*}(x) r_{\gamma}(x) s_{\mu}(x) \\ T_{A} &=\left(\delta_{\alpha \gamma} \delta_{\beta \mu}-\delta_{\alpha \mu} \delta_{\beta \gamma}\right) p_{\alpha}^{*}(x) q_{\beta}^{*}(x) r_{\gamma}(x) s_{\mu}(x).\end{aligned}
\end{equation}
These elemental tetraquark operators are combinations of two gauge-invariant quark-antiquark constituents. The individual constituents are \textit{not} mesons since they separately do not have well-defined quantum numbers. In other words, we project the entire elemental tetraquark operator onto relevant symmetry channels, rather than each individual quark-antiquark operator.

While we chose only a handful of tetraquark operators for our final analysis, we designed hundreds of operators with differing flavor structures and displacements. We tested these operators by individually adding them to a basis of single- and multi-meson operators to see if an additional level was found. Most of the operators did not yield an additional level, but we found particular operators that did. In the $\kappa$ channel, we tested the following flavor structures: $\overline s u \overline s s$, $\overline s u \overline u u$, $\overline s u \overline d u$. We found that only operators with the $\overline s u \overline s s$ flavor structure yielded an additional finite-volume state. We tested both single-site and quadruple displacements, and found operators of both types that yielded additional finite-volume states. The quadruply-displaced operators came at a higher computational cost and offered no improvements, and so were excluded from the final operator sets. In the $a_0(980)$ channel, we tested the following flavor structures: $\overline u u \overline d u$, $\overline s s \overline d u$, $\overline d u \overline d u$. We found that only operators with the $\overline u u \overline d u$ flavor structure yielded an additional finite-volume state. We only tested single-site operators in the $a_0(980)$ channel, after finding no improvement with other displacement types in the $\kappa$ channel. We also constructed operator bases that included several tetraquark operators, and found that the number of additional levels in the energy range we examined was unchanged.
\section{Lattice Spectra Results (Preliminary)}
\subsection{$\kappa$ Channel}
\begin{figure}
  \includegraphics[scale=0.6]{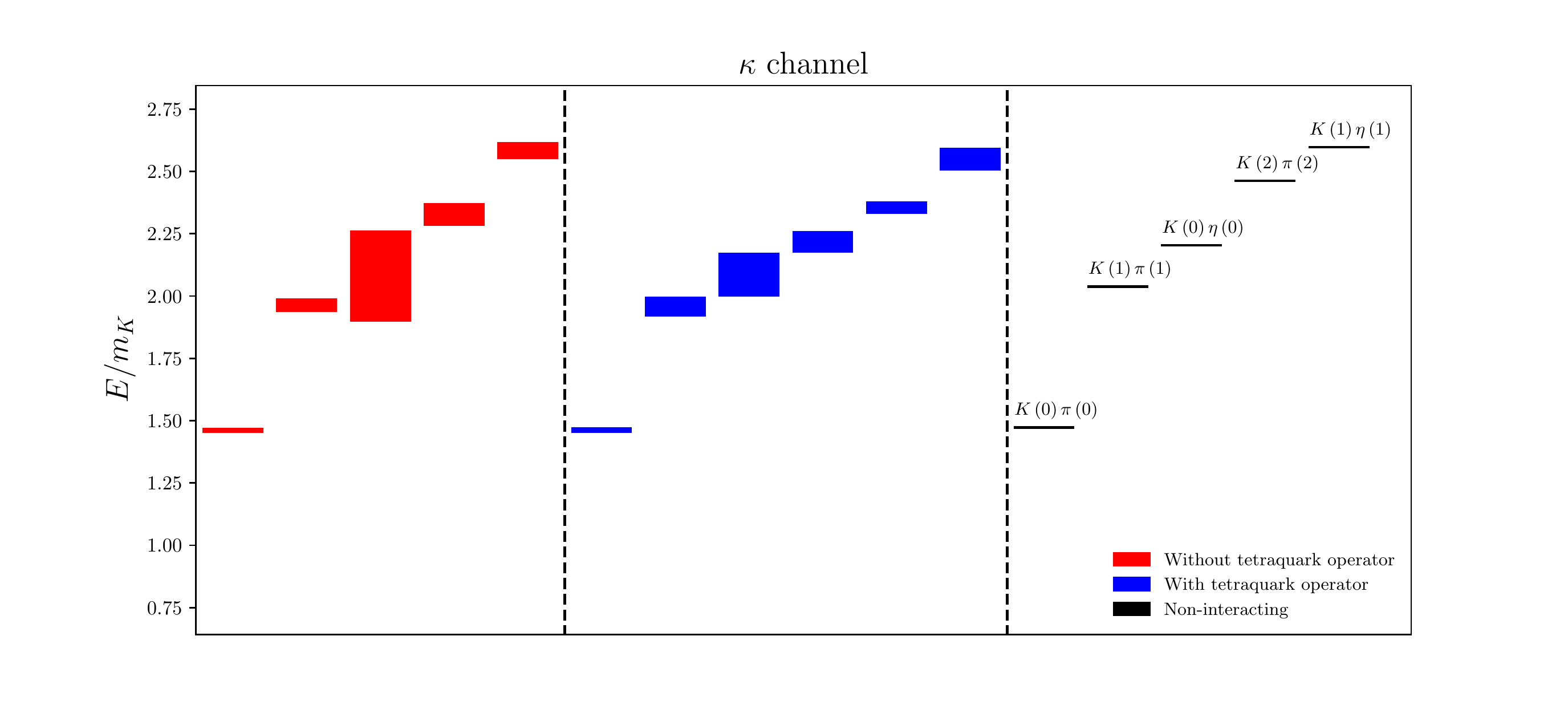}
  \caption{The first five and six levels of the spectrum in the $\kappa$ at-rest symmetry channel. On the left: the spectrum obtained using a basis with no tetraquark operators. In the middle: the spectrum obtained using one tetraquark operator. On the right: non-interacting levels shown for reference, where $(\boldsymbol{d}^2)$ denotes particles with squared momentum $(2\pi\boldsymbol{d}/L)^2$.}\label{fig:kappa_spectrum}
\end{figure}

\begin{figure}
  \includegraphics[scale=0.5]{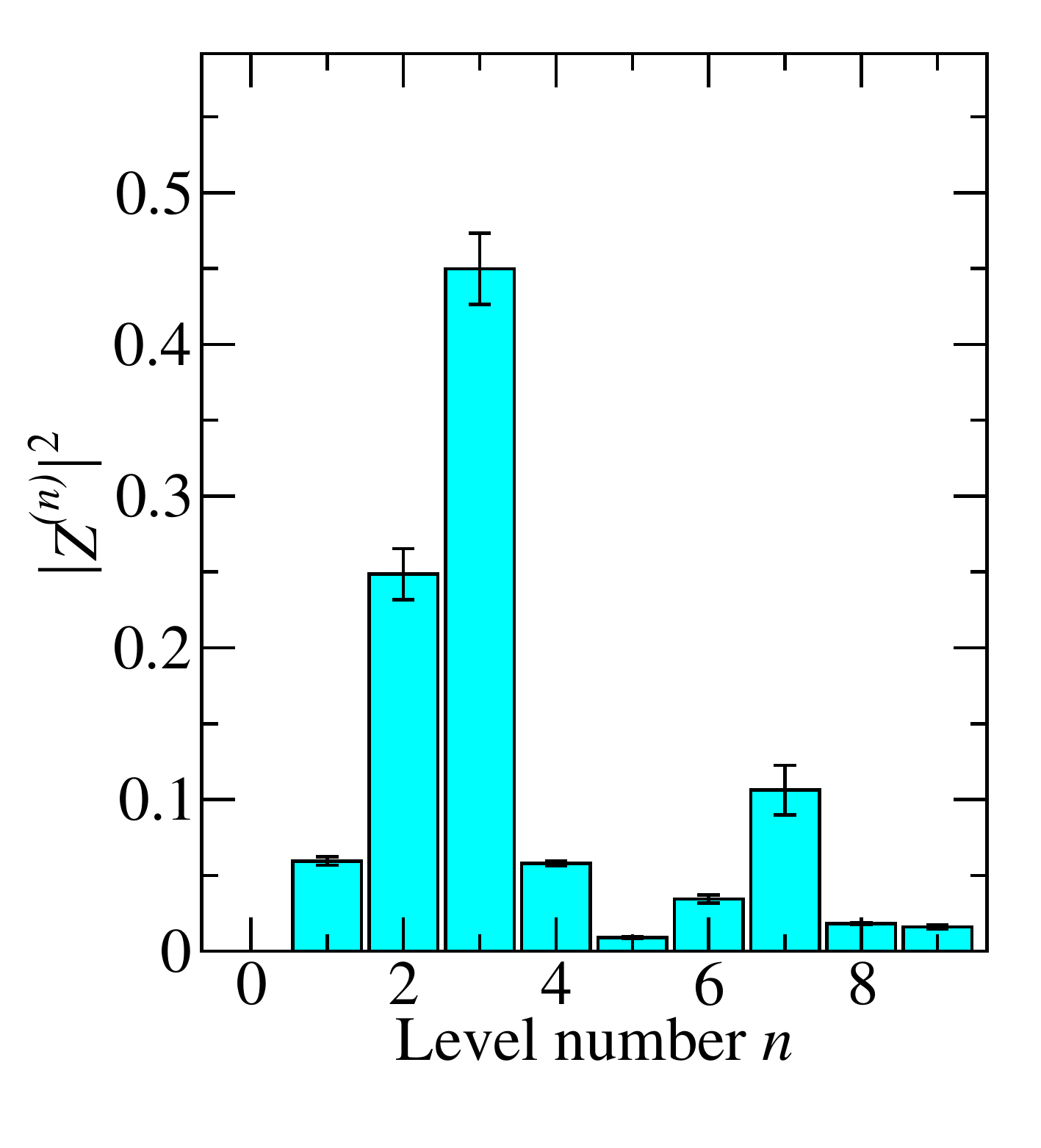}
  \caption{The overlap factors for the tetraquark operator used to produce the extra level in the $\kappa$ symmetry channel.}\label{fig:zkappa}
\end{figure}

We summarize results obtained by fitting a spectrum in the $\kappa$ at-rest symmetry channel for two operator bases: one including only single-meson and two-meson operators, and one including single-meson, two-meson, and tetraquark operators. Figure \ref{fig:kappa_spectrum} shows the spectrum with and without the inclusion of a tetraquark operator in the basis. The tetraquark operator is of the flavor structure $\overline s u \overline s s$, is of the antisymmetric form in Eq.~(\ref{eq:tsta}), and has no quark displacement. We found that single-site ($\boldsymbol{d_\alpha} = \boldsymbol{d_\beta} = 0$) tetraquark operators resulted in better (less noisy) correlator signals than displaced operators. We see that including a tetraquark operator yields an additional finite-volume state in the range of $(2.178 - 2.256) m_K$, which is not present when only single- and two-meson operators are used. Additionally, a plot of the overlap factors for the tetraquark operator (Figure~\ref{fig:zkappa}) shows significant overlap onto this extra state (level 3 in the plot). This suggests that there is a finite-volume state in our lattice spectrum that shares quantum numbers with the $\kappa$ resonance, and that has tetraquark content. Whether or not this is evidence of the $\kappa$ resonance having tetraquark content, however, will have to wait for future scattering studies using Lüscher's method.
\subsection{$a_0(980)$ Channel}
\begin{figure}
  \includegraphics[scale=0.6]{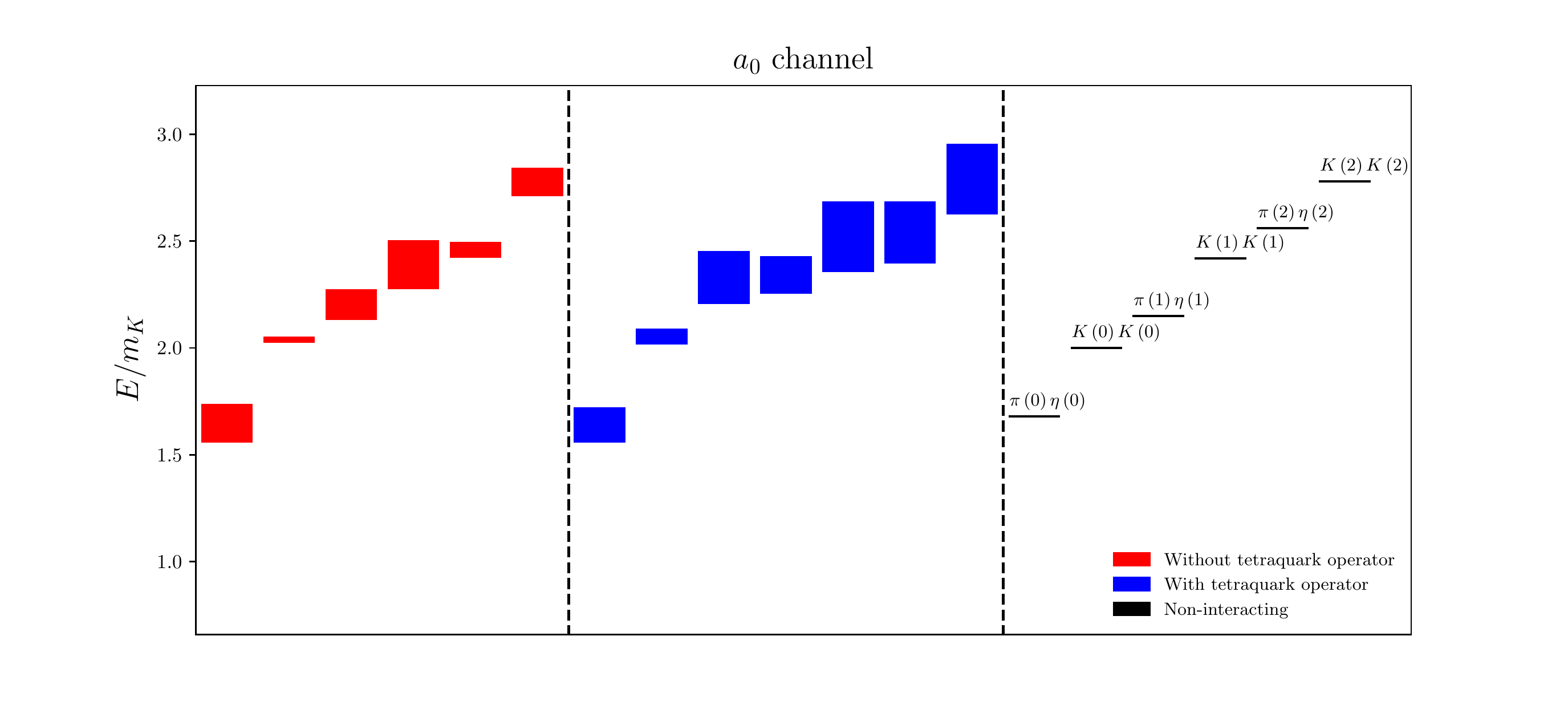}
  \caption{The first six and seven levels of the spectrum in the $a_0(980)$ at-rest symmetry channel. On the left: the spectrum obtained using a basis with no tetraquark operators. In the middle: the spectrum obtained using one tetraquark operator. On the right: non-interacting levels shown for reference, where $(\boldsymbol{d}^2)$ denotes particles with squared momentum $(2\pi\boldsymbol{d}/L)^2$.}\label{fig:a0_spectrum}
\end{figure}

\begin{figure}
  \includegraphics[scale=0.5]{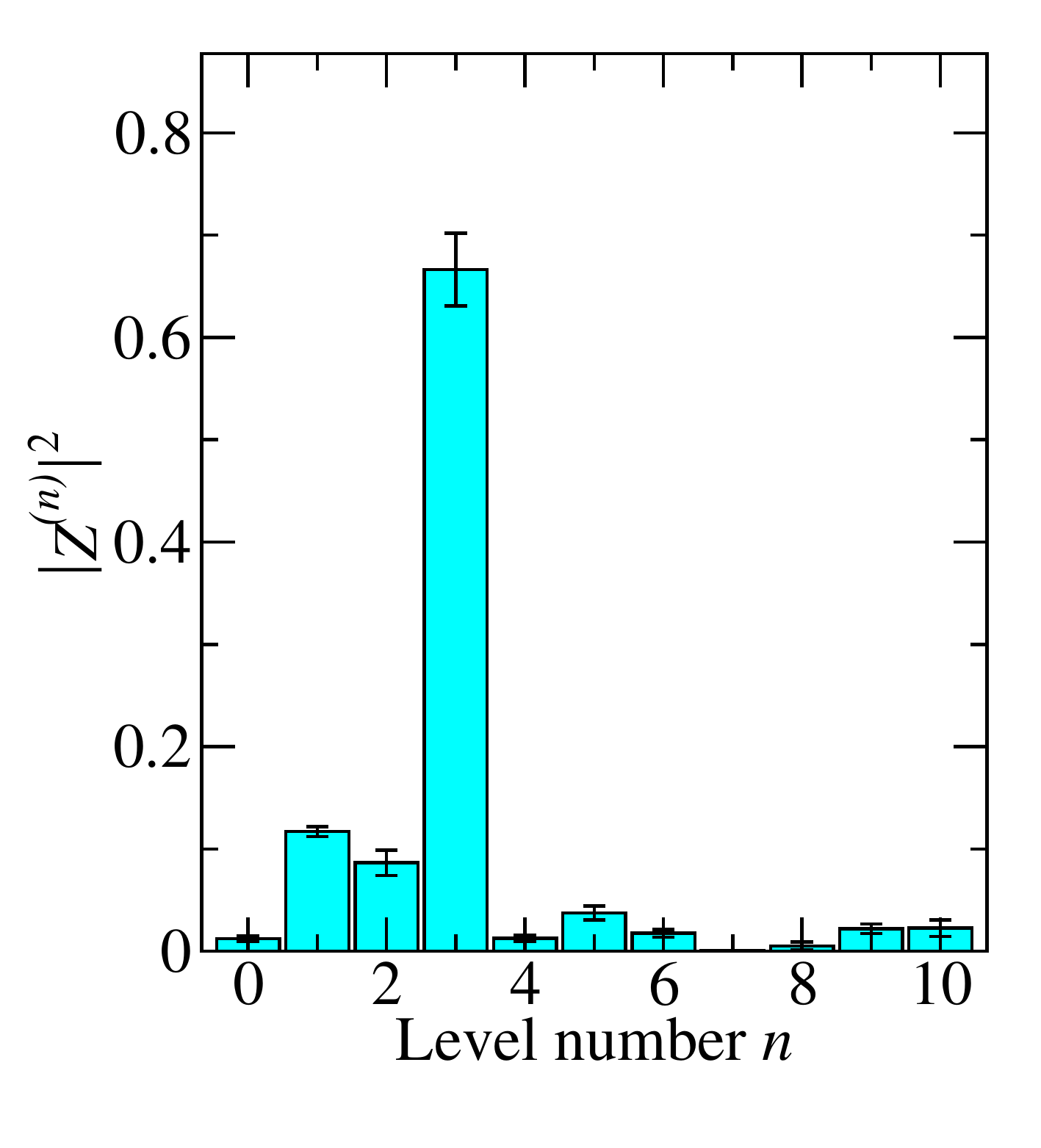}
  \caption{The overlap factors for the tetraquark operator used to produce the extra level in the $a_0(980)$ symmetry channel.}\label{fig:za0}
\end{figure}

We summarize results obtained by fitting a spectrum in the $a_0(980)$ at-rest symmetry channel for again for two operator bases as in the $\kappa$ channel. Figure \ref{fig:a0_spectrum} shows the spectrum with and without the inclusion of a tetraquark operator in the basis. The tetraquark operator is of the flavor structure $\overline u u \overline d u$, is also of the antisymmetric form in (\ref{eq:tsta}), and again has no quark displacement. We again found that using single-site tetraquark operators resulted in better correlator signals than displaced operators. We see an extra level appear in the range of $(2.258 - 2.426) m_K$ when we include a tetraquark operator. Again, overlap factors are shown for the tetraquark operator, and significant overlaps with the additional level (level 3) can be seen in Figure~\ref{fig:za0}. This suggests there is a finite-volume state in our lattice spectrum that shares quantum numbers with the $a_0(980)$ resonance, and that has tetraquark content. As in the $\kappa$-channel case, evidence for or against the $a_0(980)$ having tetraquark content will have to wait for future scattering studies done by applying Lüscher's method.
\section{Conclusions}
We have presented results detailing the effect of including tetraquark operators on determining the lattice spectrum in each of the $\kappa$ and $a_0(980)$ symmetry channels of $N_f=2+1$ QCD with $m_\pi \approx 230$ MeV. We find that the inclusion of a tetraquark operator in the $\kappa$ channel yields an additional finite-volume state in the range of $(2.178 - 2.256)m_K$, and the inclusion of a tetraquark operator in the $a_0(980)$ channel yields an additional finite-volume state in the range of $(2.258 - 2.426) m_K$. To address the issue of tetraquark content of the $\kappa$ and $a_0(980)$ resonances, future studies employing Lüscher's method will be required.
\begin{acknowledgments}
This work was supported by the U.S. National Science Foundation under award PHY-1613449. Computing resources were provided by the Extreme Science and Engineering Discovery Environment (XSEDE) under grant number TG-MCA07S017.
\end{acknowledgments}

\nocite{*}
\bibliography{thebib}

\providecommand{\noopsort}[1]{}\providecommand{\singleletter}[1]{#1}%
\begin{thebibliography}{14}%
\makeatletter
\providecommand \@ifxundefined [1]{%
 \@ifx{#1\undefined}
}%
\providecommand \@ifnum [1]{%
 \ifnum #1\expandafter \@firstoftwo
 \else \expandafter \@secondoftwo
 \fi
}%
\providecommand \@ifx [1]{%
 \ifx #1\expandafter \@firstoftwo
 \else \expandafter \@secondoftwo
 \fi
}%
\providecommand \natexlab [1]{#1}%
\providecommand \enquote  [1]{``#1''}%
\providecommand \bibnamefont  [1]{#1}%
\providecommand \bibfnamefont [1]{#1}%
\providecommand \citenamefont [1]{#1}%
\providecommand \href@noop [0]{\@secondoftwo}%
\providecommand \href [0]{\begingroup \@sanitize@url \@href}%
\providecommand \@href[1]{\@@startlink{#1}\@@href}%
\providecommand \@@href[1]{\endgroup#1\@@endlink}%
\providecommand \@sanitize@url [0]{\catcode `\\12\catcode `\$12\catcode
  `\&12\catcode `\#12\catcode `\^12\catcode `\_12\catcode `\%12\relax}%
\providecommand \@@startlink[1]{}%
\providecommand \@@endlink[0]{}%
\providecommand \url  [0]{\begingroup\@sanitize@url \@url }%
\providecommand \@url [1]{\endgroup\@href {#1}{\urlprefix }}%
\providecommand \urlprefix  [0]{URL }%
\providecommand \Eprint [0]{\href }%
\providecommand \doibase [0]{http://dx.doi.org/}%
\providecommand \selectlanguage [0]{\@gobble}%
\providecommand \bibinfo  [0]{\@secondoftwo}%
\providecommand \bibfield  [0]{\@secondoftwo}%
\providecommand \translation [1]{[#1]}%
\providecommand \BibitemOpen [0]{}%
\providecommand \bibitemStop [0]{}%
\providecommand \bibitemNoStop [0]{.\EOS\space}%
\providecommand \EOS [0]{\spacefactor3000\relax}%
\providecommand \BibitemShut  [1]{\csname bibitem#1\endcsname}%
\let\auto@bib@innerbib\@empty
\bibitem [{\citenamefont {Jaffe}(2005)}]{Jaffe:2004ph}%
  \BibitemOpen
  \bibfield  {author} {\bibinfo {author} {\bibfnamefont {R.~L.}\ \bibnamefont
  {Jaffe}},\ }\bibfield  {title} {\enquote {\bibinfo {title} {{Exotica}},}\
  }\bibfield  {booktitle} {\emph {\bibinfo {booktitle} {{Proceedings, 6th
  International Conference on Hyperons, charm and beauty hadrons (BEACH 2004):
  Chicago, USA, June 27-July 3, 2004}}},\ }\href {\doibase
  10.1016/j.physrep.2004.11.005} {\bibfield  {journal} {\bibinfo  {journal}
  {Phys. Rept.}\ }\textbf {\bibinfo {volume} {409}},\ \bibinfo {pages} {1--45}
  (\bibinfo {year} {2005})},\ \bibinfo {note} {[,191(2004)]},\ \Eprint
  {http://arxiv.org/abs/hep-ph/0409065} {arXiv:hep-ph/0409065 [hep-ph]}
  \BibitemShut {NoStop}%
\bibitem [{\citenamefont {Amsler}\ and\ \citenamefont
  {Törnqvist}(2004)}]{AMSLER200461}%
  \BibitemOpen
  \bibfield  {author} {\bibinfo {author} {\bibfnamefont {C.}~\bibnamefont
  {Amsler}}\ and\ \bibinfo {author} {\bibfnamefont {N.~A.}\ \bibnamefont
  {Törnqvist}},\ }\bibfield  {title} {\enquote {\bibinfo {title} {Mesons
  beyond the naive quark model},}\ }\href {\doibase
  https://doi.org/10.1016/j.physrep.2003.09.003} {\bibfield  {journal}
  {\bibinfo  {journal} {Physics Reports}\ }\textbf {\bibinfo {volume} {389}},\
  \bibinfo {pages} {61 -- 117} (\bibinfo {year} {2004})}\BibitemShut {NoStop}%
\bibitem [{\citenamefont {Close}\ and\ \citenamefont
  {Törnqvist}(2002)}]{Close_2002}%
  \BibitemOpen
  \bibfield  {author} {\bibinfo {author} {\bibfnamefont {F.~E.}\ \bibnamefont
  {Close}}\ and\ \bibinfo {author} {\bibfnamefont {N.~A.}\ \bibnamefont
  {Törnqvist}},\ }\bibfield  {title} {\enquote {\bibinfo {title} {Scalar
  mesons above and below 1 {GeV}},}\ }\href {\doibase
  10.1088/0954-3899/28/10/201} {\bibfield  {journal} {\bibinfo  {journal}
  {Journal of Physics G: Nuclear and Particle Physics}\ }\textbf {\bibinfo
  {volume} {28}},\ \bibinfo {pages} {R249--R267} (\bibinfo {year}
  {2002})}\BibitemShut {NoStop}%
\bibitem [{\citenamefont {Maiani}\ \emph {et~al.}(2004)\citenamefont {Maiani},
  \citenamefont {Piccinini}, \citenamefont {Polosa},\ and\ \citenamefont
  {Riquer}}]{PhysRevLett.93.212002}%
  \BibitemOpen
  \bibfield  {author} {\bibinfo {author} {\bibfnamefont {L.}~\bibnamefont
  {Maiani}}, \bibinfo {author} {\bibfnamefont {F.}~\bibnamefont {Piccinini}},
  \bibinfo {author} {\bibfnamefont {A.~D.}\ \bibnamefont {Polosa}}, \ and\
  \bibinfo {author} {\bibfnamefont {V.}~\bibnamefont {Riquer}},\ }\bibfield
  {title} {\enquote {\bibinfo {title} {New look at scalar mesons},}\ }\href
  {\doibase 10.1103/PhysRevLett.93.212002} {\bibfield  {journal} {\bibinfo
  {journal} {Phys. Rev. Lett.}\ }\textbf {\bibinfo {volume} {93}},\ \bibinfo
  {pages} {212002} (\bibinfo {year} {2004})}\BibitemShut {NoStop}%
\bibitem [{\citenamefont {Prelovsek}\ \emph {et~al.}(2010)\citenamefont
  {Prelovsek}, \citenamefont {Draper}, \citenamefont {Lang}, \citenamefont
  {Limmer}, \citenamefont {Liu}, \citenamefont {Mathur},\ and\ \citenamefont
  {Mohler}}]{Prelovsek2010}%
  \BibitemOpen
  \bibfield  {author} {\bibinfo {author} {\bibfnamefont {S.}~\bibnamefont
  {Prelovsek}}, \bibinfo {author} {\bibfnamefont {T.}~\bibnamefont {Draper}},
  \bibinfo {author} {\bibfnamefont {C.~B.}\ \bibnamefont {Lang}}, \bibinfo
  {author} {\bibfnamefont {M.}~\bibnamefont {Limmer}}, \bibinfo {author}
  {\bibfnamefont {K.-F.}\ \bibnamefont {Liu}}, \bibinfo {author} {\bibfnamefont
  {N.}~\bibnamefont {Mathur}}, \ and\ \bibinfo {author} {\bibfnamefont
  {D.}~\bibnamefont {Mohler}},\ }\bibfield  {title} {\enquote {\bibinfo {title}
  {{Lattice study of light scalar tetraquarks with $I = 0, 2, \frac{1}{2},
  \frac{3}{2}$: Are $\sigma$ and $\kappa$ tetraquarks?}}}\ }\href {\doibase
  10.1103/PhysRevD.82.094507} {\bibfield  {journal} {\bibinfo  {journal}
  {Physical Review D}\ }\textbf {\bibinfo {volume} {82}},\ \bibinfo {pages}
  {094507} (\bibinfo {year} {2010})}\BibitemShut {NoStop}%
\bibitem [{\citenamefont {Alexandrou}\ \emph {et~al.}(2013)\citenamefont
  {Alexandrou}, \citenamefont {Daldrop}, \citenamefont {{Dalla Brida}},
  \citenamefont {Gravina}, \citenamefont {Scorzato}, \citenamefont {Urbach},\
  and\ \citenamefont {Wagner}}]{Alexandrou2013}%
  \BibitemOpen
  \bibfield  {author} {\bibinfo {author} {\bibfnamefont {C.}~\bibnamefont
  {Alexandrou}}, \bibinfo {author} {\bibfnamefont {J.~O.}\ \bibnamefont
  {Daldrop}}, \bibinfo {author} {\bibfnamefont {M.}~\bibnamefont {{Dalla
  Brida}}}, \bibinfo {author} {\bibfnamefont {M.}~\bibnamefont {Gravina}},
  \bibinfo {author} {\bibfnamefont {L.}~\bibnamefont {Scorzato}}, \bibinfo
  {author} {\bibfnamefont {C.}~\bibnamefont {Urbach}}, \ and\ \bibinfo {author}
  {\bibfnamefont {M.}~\bibnamefont {Wagner}},\ }\bibfield  {title} {\enquote
  {\bibinfo {title} {{Lattice investigation of the scalar mesons $a_0(980)$ and
  $\kappa$ using four-quark operators}},}\ }\href {\doibase
  10.1007/JHEP04(2013)137} {\bibfield  {journal} {\bibinfo  {journal} {JHEP04}\
  ,\ \bibinfo {pages} {137}} (\bibinfo {year} {2013})}\BibitemShut {NoStop}%
\bibitem [{\citenamefont {Alexandrou}\ \emph {et~al.}(2018)\citenamefont
  {Alexandrou}, \citenamefont {Berlin}, \citenamefont {{Dalla Brida}},
  \citenamefont {Finkenrath}, \citenamefont {Leontiou},\ and\ \citenamefont
  {Wagner}}]{Alexandrou2018}%
  \BibitemOpen
  \bibfield  {author} {\bibinfo {author} {\bibfnamefont {C.}~\bibnamefont
  {Alexandrou}}, \bibinfo {author} {\bibfnamefont {J.}~\bibnamefont {Berlin}},
  \bibinfo {author} {\bibfnamefont {M.}~\bibnamefont {{Dalla Brida}}}, \bibinfo
  {author} {\bibfnamefont {J.}~\bibnamefont {Finkenrath}}, \bibinfo {author}
  {\bibfnamefont {T.}~\bibnamefont {Leontiou}}, \ and\ \bibinfo {author}
  {\bibfnamefont {M.}~\bibnamefont {Wagner}},\ }\bibfield  {title} {\enquote
  {\bibinfo {title} {{Lattice QCD investigation of the structure of the
  $a_0(980)$ meson}},}\ }\href {\doibase 10.1103/PhysRevD.97.034506} {\bibfield
   {journal} {\bibinfo  {journal} {Physical Review D}\ }\textbf {\bibinfo
  {volume} {97}} (\bibinfo {year} {2018}),\
  10.1103/PhysRevD.97.034506}\BibitemShut {NoStop}%
\bibitem [{\citenamefont {Morningstar}\ \emph {et~al.}(2011)\citenamefont
  {Morningstar}, \citenamefont {Bulava}, \citenamefont {Foley}, \citenamefont
  {Juge}, \citenamefont {Lenkner}, \citenamefont {Peardon},\ and\ \citenamefont
  {Wong}}]{slaph}%
  \BibitemOpen
  \bibfield  {author} {\bibinfo {author} {\bibfnamefont {C.}~\bibnamefont
  {Morningstar}}, \bibinfo {author} {\bibfnamefont {J.}~\bibnamefont {Bulava}},
  \bibinfo {author} {\bibfnamefont {J.}~\bibnamefont {Foley}}, \bibinfo
  {author} {\bibfnamefont {K.~J.}\ \bibnamefont {Juge}}, \bibinfo {author}
  {\bibfnamefont {D.}~\bibnamefont {Lenkner}}, \bibinfo {author} {\bibfnamefont
  {M.}~\bibnamefont {Peardon}}, \ and\ \bibinfo {author} {\bibfnamefont
  {C.~H.}\ \bibnamefont {Wong}},\ }\bibfield  {title} {\enquote {\bibinfo
  {title} {{Improved stochastic estimation of quark propagation with Laplacian
  Heaviside smearing in lattice QCD}},}\ }\href {\doibase
  10.1103/PhysRevD.83.114505} {\bibfield  {journal} {\bibinfo  {journal} {Phys.
  Rev.}\ }\textbf {\bibinfo {volume} {D83}},\ \bibinfo {pages} {114505}
  (\bibinfo {year} {2011})},\ \Eprint {http://arxiv.org/abs/1104.3870}
  {arXiv:1104.3870 [hep-lat]} \BibitemShut {NoStop}%
\bibitem [{\citenamefont {Edwards}, \citenamefont {Jo\'o},\ and\ \citenamefont
  {Lin}(2008)}]{PhysRevD.78.054501}%
  \BibitemOpen
  \bibfield  {author} {\bibinfo {author} {\bibfnamefont {R.~G.}\ \bibnamefont
  {Edwards}}, \bibinfo {author} {\bibfnamefont {B.}~\bibnamefont {Jo\'o}}, \
  and\ \bibinfo {author} {\bibfnamefont {H.-W.}\ \bibnamefont {Lin}},\
  }\bibfield  {title} {\enquote {\bibinfo {title} {Tuning for three flavors of
  anisotropic clover fermions with stout-link smearing},}\ }\href {\doibase
  10.1103/PhysRevD.78.054501} {\bibfield  {journal} {\bibinfo  {journal} {Phys.
  Rev. D}\ }\textbf {\bibinfo {volume} {78}},\ \bibinfo {pages} {054501}
  (\bibinfo {year} {2008})}\BibitemShut {NoStop}%
\bibitem [{\citenamefont {Bulava}\ \emph {et~al.}(2009)\citenamefont {Bulava},
  \citenamefont {Edwards}, \citenamefont {Engelson}, \citenamefont {Foley},
  \citenamefont {Jo\'o}, \citenamefont {Lichtl}, \citenamefont {Lin},
  \citenamefont {Mathur}, \citenamefont {Morningstar}, \citenamefont
  {Richards},\ and\ \citenamefont {Wallace}}]{PhysRevD.79.034505}%
  \BibitemOpen
  \bibfield  {author} {\bibinfo {author} {\bibfnamefont {J.~M.}\ \bibnamefont
  {Bulava}}, \bibinfo {author} {\bibfnamefont {R.~G.}\ \bibnamefont {Edwards}},
  \bibinfo {author} {\bibfnamefont {E.}~\bibnamefont {Engelson}}, \bibinfo
  {author} {\bibfnamefont {J.}~\bibnamefont {Foley}}, \bibinfo {author}
  {\bibfnamefont {B.}~\bibnamefont {Jo\'o}}, \bibinfo {author} {\bibfnamefont
  {A.}~\bibnamefont {Lichtl}}, \bibinfo {author} {\bibfnamefont {H.-W.}\
  \bibnamefont {Lin}}, \bibinfo {author} {\bibfnamefont {N.}~\bibnamefont
  {Mathur}}, \bibinfo {author} {\bibfnamefont {C.}~\bibnamefont {Morningstar}},
  \bibinfo {author} {\bibfnamefont {D.~G.}\ \bibnamefont {Richards}}, \ and\
  \bibinfo {author} {\bibfnamefont {S.~J.}\ \bibnamefont {Wallace}},\
  }\bibfield  {title} {\enquote {\bibinfo {title} {Excited state nucleon
  spectrum with two flavors of dynamical fermions},}\ }\href {\doibase
  10.1103/PhysRevD.79.034505} {\bibfield  {journal} {\bibinfo  {journal} {Phys.
  Rev. D}\ }\textbf {\bibinfo {volume} {79}},\ \bibinfo {pages} {034505}
  (\bibinfo {year} {2009})}\BibitemShut {NoStop}%
\bibitem [{\citenamefont {Clark}, \citenamefont {Kennedy},\ and\ \citenamefont
  {Sroczynski}(2005)}]{CLARK2005835}%
  \BibitemOpen
  \bibfield  {author} {\bibinfo {author} {\bibfnamefont {M.}~\bibnamefont
  {Clark}}, \bibinfo {author} {\bibfnamefont {A.}~\bibnamefont {Kennedy}}, \
  and\ \bibinfo {author} {\bibfnamefont {Z.}~\bibnamefont {Sroczynski}},\
  }\bibfield  {title} {\enquote {\bibinfo {title} {Exact 2+1 flavour {RHMC}
  simulations},}\ }\href {\doibase
  https://doi.org/10.1016/j.nuclphysbps.2004.11.192} {\bibfield  {journal}
  {\bibinfo  {journal} {Nuclear Physics B - Proceedings Supplements}\ }\textbf
  {\bibinfo {volume} {140}},\ \bibinfo {pages} {835 -- 837} (\bibinfo {year}
  {2005})},\ \bibinfo {note} {{LATTICE} 2004}\BibitemShut {NoStop}%
\bibitem [{\citenamefont {Lüscher}\ and\ \citenamefont
  {Wolff}(1990)}]{LUSCHER1990222}%
  \BibitemOpen
  \bibfield  {author} {\bibinfo {author} {\bibfnamefont {M.}~\bibnamefont
  {Lüscher}}\ and\ \bibinfo {author} {\bibfnamefont {U.}~\bibnamefont
  {Wolff}},\ }\bibfield  {title} {\enquote {\bibinfo {title} {How to calculate
  the elastic scattering matrix in two-dimensional quantum field theories by
  numerical simulation},}\ }\href {\doibase
  https://doi.org/10.1016/0550-3213(90)90540-T} {\bibfield  {journal} {\bibinfo
   {journal} {Nuclear Physics B}\ }\textbf {\bibinfo {volume} {339}},\ \bibinfo
  {pages} {222 -- 252} (\bibinfo {year} {1990})}\BibitemShut {NoStop}%
\bibitem [{\citenamefont {Morningstar}\ and\ \citenamefont
  {Peardon}(2004)}]{PhysRevD.69.054501}%
  \BibitemOpen
  \bibfield  {author} {\bibinfo {author} {\bibfnamefont {C.}~\bibnamefont
  {Morningstar}}\ and\ \bibinfo {author} {\bibfnamefont {M.}~\bibnamefont
  {Peardon}},\ }\bibfield  {title} {\enquote {\bibinfo {title} {Analytic
  smearing of $\mathrm{SU}(3)$ link variables in lattice {QCD}},}\ }\href
  {\doibase 10.1103/PhysRevD.69.054501} {\bibfield  {journal} {\bibinfo
  {journal} {Phys. Rev. D}\ }\textbf {\bibinfo {volume} {69}},\ \bibinfo
  {pages} {054501} (\bibinfo {year} {2004})}\BibitemShut {NoStop}%
\bibitem [{\citenamefont {Hanlon}(2018)}]{pittir33243}%
  \BibitemOpen
  \bibfield  {author} {\bibinfo {author} {\bibfnamefont {A.~D.}\ \bibnamefont
  {Hanlon}},\ }\href {http://d-scholarship.pitt.edu/33243/} {\enquote {\bibinfo
  {title} {The {\ensuremath{\rho}} meson spectrum and k{\ensuremath{\pi}}
  scattering with partial wave mixing in lattice {QCD}},}\ } (\bibinfo {year}
  {2018})\BibitemShut {NoStop}%
\end{thebibliography}%

\end{document}